\documentclass[prc,superscriptaddress,unsortedaddress,twocolumn,preprintnumbers,amsmath,amssymb,floatfix,dvipdfmx]{revtex4}
\usepackage{amsmath}
\usepackage{amssymb}
\usepackage{mathrsfs}
\usepackage{multirow}
\usepackage{bm}
\usepackage{ulem}
\usepackage{color}
\usepackage[dvipdfmx]{graphicx}

\newcommand{\vR}{\bm{R}}
\renewcommand{\vr}{\bm{r}}
\newcommand{\vs}{\bm{s}}

\def\mbold#1{\mbox{\boldmath $#1$}}

\newcommand{\rP}{\rm P}
\newcommand{\rT}{\rm T}

\begin{document}

\title{
Microscopic optical potentials including breakup effects\\ 
for elastic scattering 
}

\author{Shoya Ogawa}
\email[]{s-ogawa@phys.kyushu-u.ac.jp}
\affiliation{Department of Physics, Kyushu University, Fukuoka 819-0395, Japan}

\author{Ryo Horinouchi}
\email[]{horinouchi@phys.kyushu-u.ac.jp}
\affiliation{Department of Physics, Kyushu University, Fukuoka 819-0395, Japan}

\author{Masakazu Toyokawa}
\email[]{toyokawa@phys.kyushu-u.ac.jp}
\affiliation{Department of Physics, Kyushu University, Fukuoka 819-0395, Japan}

\author{Takuma Matsumoto}
\email[]{matsumoto@phys.kyushu-u.ac.jp}
\affiliation{Department of Physics, Kyushu University, Fukuoka 819-0395, Japan}

\date{\today}

\begin{abstract} 
 We construct a microscopic optical potential including breakup effects
 for elastic scattering of weakly-binding projectiles within the Glauber
 model, in which a nucleon-nucleus potential is derived by the
 $g$-matrix folding model.  
 The derived microscopic optical potential is referred to as the eikonal 
 potential. 
 For $d$ scattering, the calculation with the eikonal potential reasonably 
 reproduces the result with an exact calculation estimated by the 
 continuum-discretized coupled-channels method. 
 As the properties of the eikonal potential, the inaccuracy of the 
 eikonal approximation used in the Glauber model is partially excluded.
 We also analyse the $^6$He scattering from $^{12}$C with the eikonal
 potential and show its applicability to the scattering with many-body
 projectiles.  
\end{abstract}


\maketitle

\section{Introduction}
\label{Introduction}

Microscopic understanding of nucleon-nucleus (NA) and nucleus-nucleus
(AA) optical potentials is one of the most important issues in nuclear
reaction theory. The optical potentials are necessary to describe not only
elastic scattering but also reactions including higher-order
processes. For example, the distorted-wave Born approximation and the
continuum-discretized coupled-channels method
(CDCC)~\cite{Kam89,Aus87,Yah12} require the optical 
potentials between constructs of a projectile (P) and a target (T) to
describe inelastic scattering, breakup and transfer reactions. 

The $g$-matrix folding model has been widely used as a reliable method
to describe the optical potential~\cite{Are95,Min09,Sin75,Kho07,Fur10,Sum12}. 
The $g$ matrix is an effective nucleon-nucleon 
interaction~\cite{Ber77,Jeu77,Bri77,Sat79,Sat83,Yam83,Rik84,Amo00,Sal02,Fur09,Toy15}
in nuclear matter, and depends on the density $\rho_{\rm m}$ of the nuclear matter.    
In the $g$-matrix folding model, the optical potentials are derived by
folding the $g$-matrix with the target density $\rho_{\rm T}$ for NA
scattering, and with $\rho_{\rm T}$ and the projectile one 
$\rho_{\rm P}$ for AA scattering. The procedures are referred to as the
single-folding (SF) model for NA scattering and double-folding (DF)
model for AA scattering, respectively. 

In the SF model, 
$\rho_{\rm T}$ is referred as $\rho_{\rm m}$ in the $g$ matrix 
with the local density approximation, 
and the calculated NA optical potential becomes 
nonlocal by taking into account knock-on exchange processes. 
The nonlocal potential is not practical in many applications, 
but can be localized by the 
Brieva-Rook approximation~\cite{Bri77} in good accuracy. The localized
potential is quite successful in reproducing experimental data
systematically, when reliable $g$ matrices such as
Melbourne~\cite{Amo00}, CEG~\cite{Yam83,Fur09}, and
$\chi$EFT~\cite{Toy15} $g$ matrices are adopted.  

On the other hand, it is more difficult to derivate the optical
potential with the DF model for AA scattering than that with the SF model for
NA scattering. The main problem of the DF model is how to treat
$\rho_{\rm m}$ in the $g$ matrix for AA scattering. In general, the
frozen-density approximation (FDA) is applied, in which the sum of
$\rho_{\rm P}$ and $\rho_{\rm T}$ is taken as $\rho_{\rm m}$ in the $g$
matrix. The optical potential derived by the DF model with the FDA
often needs a normalization factor for the real and imaginary parts 
to reproduce experimental data. Thus the choice of $\rho_m$ in 
the $g$ matrix for the DF model is a longstanding open problem.

In the previous works for $^{3,4}$He scattering on various
targets~\cite{Ega14,Toy15b}, we proposed the target-density
approximation (TDA) in the DF model, where $\rho_{\rm m}$ is estimated
with only $\rho_{\rm T}$. 
The DF model with the TDA (the DF-TDA model)
well reproduces experimental data for $^{3,4}$He scattering with no
adjustable parameter, and its theoretical validity 
can be confirmed by using multiple scattering theory~\cite{Wat53,Ker59,Yah08}. 
Furthermore, as a practical model of the DF-TDA model, we also proposed the
double-single folding (DSF) model for $^{3,4}$He scattering, in which
the optical potential between $^{3,4}$He and the target is derived by
folding the localized NA optical potential of the SF model with the
$^{3,4}$He densities. 

However, the optical potential calculated by the DSF model is not taken 
into account breakup effects on the elastic scattering.
In fact, results with the DSF model are slightly different from the 
experimental data for $^3$He scattering at low incident 
energies~\cite{Toy15b}, 
where breakup effects are important.
The discrepancy can be solved by applying the CDCC calculation
that can describe breakup effects accurately.
CDCC has been successful in analyses of
scattering of weakly-binding two- and three-body projectiles.
Although CDCC is a reliable method to analyze breakup reactions,
it is hard to apply CDCC to scattering of more than four-body projectiles 
such as $^{8}$He, because of a high computational cost.

As another approach to describe excitation effects of P and T, the
Glauber model~\cite{Gla59} has been applied to analyses of AA scattering at
intermediate energies. 
In the Glauber model, nucleon degrees of
freedom of P and T are treated by the adiabatic approximation and the
eikonal approximation. 
One of advantages of the Glauber model is to construct 
the optical potential including multiple-scattering effects from
the phase-shift
function~\cite{Gla59,Yab92,Abu00}. The derived optical potential is
called the eikonal potential. Although the eikonal potential has been 
used to analyses of scattering of weakly-binding projectiles
in the previous works~\cite{Yab92,Abu00}, the validity of the eikonal
potential for incident energies and its property have never been
clarified.  

In this paper, we investigate the validity of the eikonal potential
compared with results of CDCC with the SF model for $d$ scattering at
20--200 MeV/nucleon, which is described by a $p$ + $n$ + T three-body
model. The optical potentials for $p$-T and $n$-T systems are derived by
the SF model with the Melbourne $g$ matrix. 
Furthermore, in order to discuss the applicability of the eikonal
potential to many-body systems, we adopt the eikonal potential to
analyses of $^6$He scattering that is described by a $^4$He + $n$ + $n$
+ T four-body model.  

The paper is organized as follows. In Sec. II, we describe the
theoretical framework to derive the eikonal potential for $d$
scattering. In Sec. III, numerical results are shown, and properties of
the eikonal potential are discussed. Finally, we give a summary in Sec. IV.

\section{Theoretical framework}
\label{Theoretical framework}

\subsection{Glauber model}
\label{Theor:Glauber}

We consider $d$ scattering on T by a $p$ + $n$ + T three-body model.
Note that the below formulation can be easily extend to a
three-body projectile system such as scattering of $^6$He described by
a $^4$He + $n$ + $n$ + T four-body model.
The scattering is described by the three-body Schr\"odinger equation   
\begin{eqnarray}
 \left[
  -\frac{\hbar^2}{2\mu}\bm{\nabla}_{\vR}^2
  +U(\vr,\vR)+h_d-E
 \right]\Psi(\vr,\vR)=0,
\label{eq:Sch-many}
\end{eqnarray}
where $\vR$ and $\mu$ are the relative coordinate and the reduced mass
of the $d$-T system, respectively. 
The potential $U$ between $d$ and T is represented by
\begin{eqnarray}
 U(\vr,\vR)=U_{n}(\vr,\vR)+U_{p}(\vr,\vR)+V_{\rm C}(R),
\end{eqnarray}
where $U_{n}$ ($U_{p}$) is the optical potential between $n$ ($p$)
and T. In the present analysis, we neglect Coulomb breakup processes,
and the the Coulomb potential $V_{\rm C}$ thus depends on only $R$.
As the internal Hamiltonian $h_d$ for $d$, we adopt a simple form
with the Ohmura potential between $p$ and $n$~\cite{Ohm77}.

Initially, $d$ is the ground state, and has 
$\hbar\bm{K}$ as the relative momentum according to $R$ at $Z=-\infty$,
where $Z$ is the $z$-axis component of $\vR$ , and $\bm{K}$ is set
to be parallel to $Z$. Under this initial condition,
$\Psi$ is represented by
\begin{eqnarray}
 \Psi(\vr,\vR)
  \xrightarrow[Z\to-\infty]{}
  e^{iKZ+\cdots}\Phi_0(\vr),
\end{eqnarray}
where the ``$\cdots$'' represents effects of the Coulomb distortion.
The ground state wave function $\Phi_0$ of $d$ satisfies 
\begin{eqnarray}
 h_d\Phi_0(\vr)&=&\epsilon_0\Phi_0(\vr),
\end{eqnarray}
and the total energy conservation is defined as
\begin{eqnarray}
 \frac{\hbar^2K^2}{2\mu}=E-\epsilon_0\equiv E_0.
\end{eqnarray}

The Glauber model is based on the adiabatic approximation and the
eikonal approximation. In the adiabatic approximation, $h_d$ in
Eq.~\eqref{eq:Sch-many} is 
replaced by $\epsilon_0$. In the eikonal approximation, $\Psi$ is
described as the product of a plane wave by a new function $\hat{\Psi}$, 
\begin{eqnarray}
 \Psi(\vr,\vR)&=&e^{iKZ}\hat{\Psi}(\vr,\vR). \label{eq:Eik-wf}
\end{eqnarray}
Inserting Eq.~\eqref{eq:Eik-wf} into Eq.~\eqref{eq:Sch-many} with the
adiabatic approximation, the equation for $\hat{\Psi}$ is obtained as
\begin{eqnarray}
 \left[
 -\frac{\hbar^2}{2\mu}\bm{\nabla}_{\vR}^2
  -i\hbar v\frac{\partial}{\partial Z}
  +U(\vr,\vR)
 \right]\hat{\Psi}(\vr,\vR)=0,\label{eq:Eik-eq1}
\end{eqnarray}
where $v=\hbar K/\mu$. As an essence of the eikonal approximation, the
second derivatives of $\hat{\Psi}$ for $\vR$ is neglected by
assuming that $\hat{\Psi}$ varies smoothly with $\vR$. Accordingly,
Eq.~\eqref{eq:Eik-eq1} is rewritten as
\begin{eqnarray}
 i\hbar v\frac{\partial}{\partial Z}\hat{\Psi}(\vr,\vR)
  &=&U(\vr,\vR)\hat{\Psi}(\vr,\vR).\label{eq:Eik-eq2}
\end{eqnarray}
The conditions for the accuracy of the eikonal approximation are
well known as
\begin{eqnarray}
\frac{\left|U\right|}{E_0}<<1,\quad  a_UK>>1,
\label{eq:Eik-cond}
\end{eqnarray}
where $a_U$ is the radius of $U$.

If $U$ does not include $V_{\rm C}$, Eq.~\eqref{eq:Eik-eq2} can be
solved analytically under the initial condition as 
\begin{eqnarray}
 \hat{\Psi}(\vr,\vR)=
  \exp\left[
       -\frac{i}{\hbar v}\int_{-\infty}^Z
       U(\vr,\bm{b},Z')dZ'
      \right]\Phi_0(\vr),
\end{eqnarray}
where $\bm{b}$ is the projection of $\vR$ onto the $x$-$y$ plane.
In the eikonal approximation, the elastic scattering amplitude is 
given by
\begin{eqnarray}
 f_{\rm el}(\bm{q})=
  \frac{iK}{2\pi}
  \int d\bm{b}e^{-i\bm{q}\cdot\bm{b}}
  \left[
   1-{\cal S}(\bm{b})
  \right],
\end{eqnarray}
where the forward-scattering approximation is adopted, and $\bm{q}$ is
the transfer momentum.
The  eikonal $S$ matrix, ${\cal S}$, is 
defined as 
\begin{eqnarray}
{\cal S}(\bm{b})&=&
  \langle\Phi_0|
  e^{i\chi(\vr,\bm{b})}
  |\Phi_0\rangle
\end{eqnarray}
with the phase-shift function
\begin{eqnarray}
 \chi(\vr,\bm{b})&=&-\frac{1}{\hbar v}
  \int_{-\infty}^\infty U(\vr,\vR)dZ.
\end{eqnarray}
This expression is valid for $U$ without $V_{\rm C}$. 
When $U$ includes $V_{\rm C}$, we need a special treatment 
because of the well-known logarithmic divergence of $\chi$.

To describe scattering with the Coulomb interaction in the Glauber
model, various approaches have been proposed so far. In this study, we  
apply the simplest way with the sharp cut screened Coulomb potential
with the cut off radius $a_{\rm C}$, and 
the elastic scattering amplitude is thus
rewritten as 
\begin{eqnarray}
 f_{\rm el}^{\rm G}(\mbold{q})&=&
  \left[
   f_{\rm Ruth}(\mbold{q})
   +f_{\rm N}(\mbold{q})   
  \right]e^{-2i\eta \ln (2ka_{\rm C})},
\label{eq:fq2}
\end{eqnarray}
where $f_{\rm Ruth}$ is the the Rutherford amplitude and $\eta$ is the
Zommerfeld parameter. $f_{\rm N}$ represents the nuclear part of the
scattering amplitude given by
\begin{eqnarray}
  f_{\rm N}(\mbold{q})=
  \frac{iK}{2\pi}\int d\bm{b}\;e^{-i\bm{q}\cdot\bm{b}+2i\eta \ln (kb)}
  \left[
   1-{\cal S}_{\rm N}(\bm{b})
       \right]
\end{eqnarray}
with the nuclear part of the eikonal $S$ matrix 
\begin{eqnarray}
 {\cal S}_{\rm N}(\bm{b})=
  \langle\Phi_0|e^{i\chi_{\rm N}(\vr,\bm{b})}|\Phi_0\rangle
  \label{eq:SN}
\end{eqnarray}
and the phase-shift function 
\begin{eqnarray}
  \chi_{\rm N}(\vr,\bm{b})=
  -\frac{1}{\hbar v}
  \int_{-\infty}^\infty 
  \left[U_n(\vr,\vR)
   +U_p(\vr,\vR)
  \right]
  dZ.  \label{eq:chiN}
\end{eqnarray}
Here it should be noted that although $f_{\rm el}$ of
Eq.~\eqref{eq:fq2} depends on $a_{\rm C}$, the differential cross
section defined as $|f_{\rm el}^{\rm G}(\bm{q})|^2$ does not depend on
$a_{\rm C}$. In this paper, we refer to the calculation with 
$f_{\rm el}^{\rm G}(\mbold{q})$ as the Glauber model.

\begin{figure*}[htbp]
 \begin{center}
  \includegraphics[width=0.42\textwidth,clip]{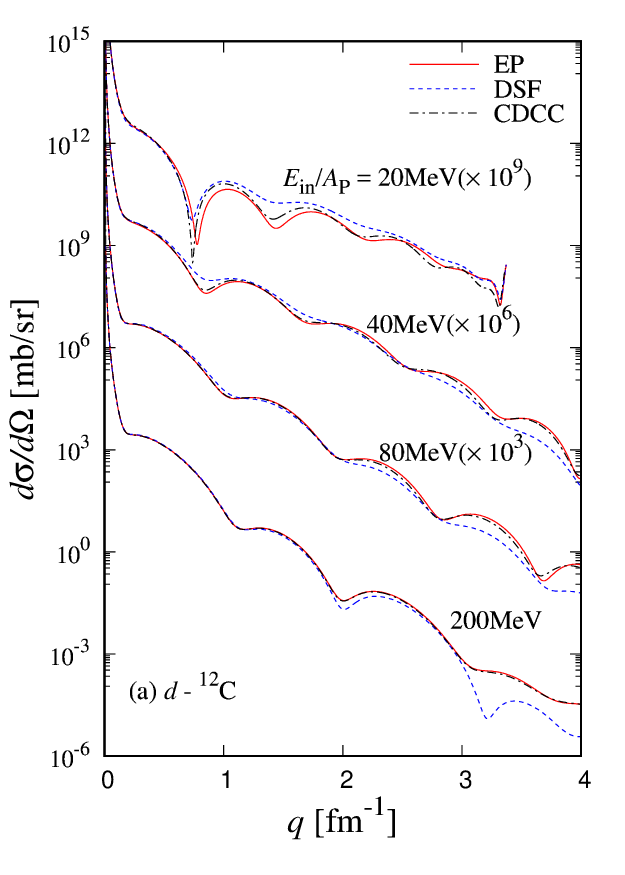}
  \includegraphics[width=0.42\textwidth,clip]{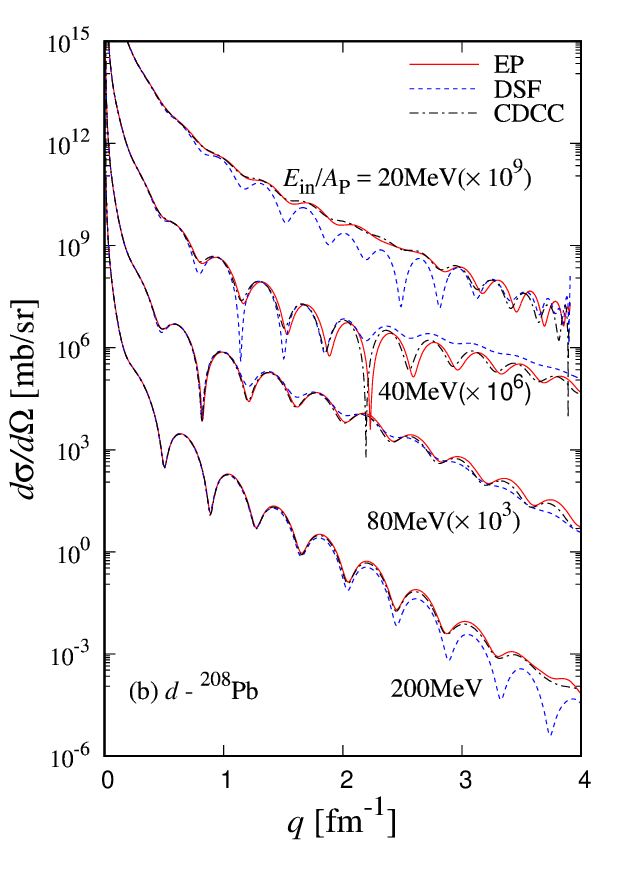}
  \caption{
  Differential cross sections $d\sigma/d\Omega$ as a function of 
  transfer momentum $q$ for $d$ scattering from (a) $^{12}$C  
  and (b) $^{208}$Pb targets at $E_{\rm in}/A_{\rP} = 20$--$200$ MeV. 
  The solid line represents the result of the EP model, 
  the dashed line denotes the result of the DSF model, 
  and the dot-dashed line corresponds to the result of CDCC, respectively. 
  Each cross section is multiplied by the factor shown in the figure. }
  \label{fig1}
 \end{center}
\end{figure*}

\subsection{Eikonal potential model}

According to the approach proposed by Glauber, we can derive the local
optical potential $U_{\rm opt}$ in the following. 
First, we introduce a new phase-shift function $\tilde{\chi}_{\rm N}$ 
to reproduce ${\cal S}_{\rm N}$ obtained from Eqs.~\eqref{eq:SN} and
\eqref{eq:chiN}; 
\begin{eqnarray}
 {\cal S}_N(b)&=&e^{i\tilde{\chi}_{\rm N}(b)}.
  \label{eq:SN1}
\end{eqnarray}
Here it should be noted that ${\cal S}_{\rm N}$ depends on only
$b=|\bm{b}|$ since $U_n$ and $U_p$ are taken into account only the central
parts as mentioned later. Then we assume that $\tilde{\chi}_{\rm N}$ is
described by the local and spherical optical potential $U_{\rm E}$ as
\begin{eqnarray}
 \tilde{\chi}_{\rm N}(b)&=&-\frac{1}{\hbar v}
  \int_{-\infty}^\infty U_{\rm E}(R)dZ. \label{eq:chiN2}
\end{eqnarray}
Finally, Solving Eq.~\eqref{eq:chiN2} for $U_{\rm E}$, we obtain the
following form
\begin{eqnarray}
 U_{\rm E}(R)&=&\frac{\hbar v}{\pi}\frac{1}{R}\frac{d}{dR}\int_{R}^{\infty}
  bdb~\frac{\tilde{\chi}_{\rm N}(b)}{\sqrt{b^2-R^2}}~.
\end{eqnarray}
The details for derivation of $U_{\rm E}$ are shown in Refs.~\cite{Gla59,Yab92,Abu00}.
In this paper, the calculation with $U_{\rm E}$ is called the eikonal
potential model (the EP model), in which the two-body Schr\"{o}dinger
equation for $d$-T scattering is defined as
\begin{eqnarray}
 \left[
  -\frac{\hbar^2}{2\mu}\bm{\nabla}_{\vR}^2
  +U_{\rm E}(R)+V_{\rm C}(R)-E_0
 \right]\Psi_{\rm EP}(\vR)=0.
\label{eq:Eik-model} 
\end{eqnarray}
Here it should be noted that $U_{\rm E}$ includes breakup effects of $d$
through the Glauber model. Meanwhile, the optical potential in the DSF
model (DSF potential) is defined as
\begin{eqnarray}
 U_{\rm DSF}(R)&=&\langle\Phi_0|U_{n}(\vr,\vR)
  +U_{p}(\vr,\vR)|\Phi_0\rangle,
\end{eqnarray}
in which breakup effects of $d$ are not taken into account.

\subsection{NA optical potential}
\label{Theor:folding}
In the present analysis, we derive $U_n$ and $U_p$ by the 
$g$-matrix folding model. In the $g$-matrix folding model, the optical
potential is obtained by folding the $g$ matrix with the target density,
and the knock-on exchange processes between the interacting two nucleons 
are considered as the dominant component of the antisymmetrization
\cite{Tan78,Aok83}.  
The folding potential is generally nonlocal because of the knock-on
exchange process, but the nonlocality is well localized by the
Brieva-Rook method~\cite{Bri77}. The folding potential is composed
of the direct (DR) and exchange (EX) terms  
\begin{eqnarray}
 U_{j}(R_j) = U^{\rm DR}_{j}(R_j)-U^{\rm EX}_{j}(R_j)~
\end{eqnarray}
where $j=n$ or $p$, and $R_{j}$ is the relative coordinate between
particle $j$ and T. Each term is written with the one-body and mixed
densities,  $\rho_{\rT}$ and $\tilde{\rho}_{\rT}$, as
\begin{eqnarray}
 U^{\rm DR}_{j} (R_j)&=& \sum_{\nu} \int d\vr_{\rT}
  g^{\rm DR}_{j\nu}(s,\rho)\rho_{\rT}^{(\nu)}(r_{\rT}),
  \\
 U^{\rm EX}_{j} (R_j)&=& \sum_{\nu} \int d\vr_{\rT}
  g^{\rm EX}_{j\nu}(s,\rho)\tilde{\rho}_{\rT}^{(\nu)}(\vr_{\rT},\vR_j)
  j_{0}(K(R_j)s). \nonumber\\
\end{eqnarray}
Here the index $\nu$ represents the $z$-component of isospin of the nucleons 
in T, $\vr_{\rT}$ is the internal coordinate of T, and
$\vs=\vr_{\rT}-\vR_j$.  
The $g$ matrix $g^{\rm DR(EX)}$ is the direct (exchange) term of the $g$
matrix among  
the considering nucleons. 
The mixed density can be described by using the one-body density with
the local Fermi gas approximation \cite{Neg72} as 
\begin{eqnarray}
 \tilde{\rho}_{\rT}(\vr_{\rT},\vR_j)
  =\rho_{\rm T}(|\vr_{\rT}-\vs/2|)\frac{3j_{1}
  (k_{\rm F}^{\rT}s)}{k_{\rm F}^{\rT}s}. 
\end{eqnarray} 

In the present study, we employ the Melbourne $g$ matrix \cite{Amo00}
constructed from Bonn-B nucleon-nucleon interaction \cite{Mac87}.  
The Melbourne $g$ matrix well reproduce nucleon elastic scattering with
and without localization of the exchange term \cite{Amo00,Toy13}.  
For $^{208}$Pb, the matter density is calculated by the spherical
Hartree-Fock method with the Gogny-D1S interaction \cite{GognyD1S} and
the spurious center-of-mass (c.m.) motion is removed with 
the standard procedure \cite{Sum12}. 
For $^{12}$C, we take the phenomenological proton-density determined from 
electron scattering \cite{phen-density}; 
here the finite-size effect of proton charge is unfolded with the standard 
procedure~\cite{Singhal}, and the neutron density is assumed to have the 
same geometry as the proton one.

\section{Results and Discussions}
\label{Results}
\subsection{differential cross section for $d$ scattering}
\label{Result:cross}

We investigate $d$ scattering from $^{12}$C and $^{208}$Pb targets at
$E_{\rm in}/A_{\rP}=$20--200 MeV by
using the DSF model, the EP model, and CDCC.
It is known that CDCC has been successful in analyses of
$d$ scattering~\cite{Kam89},
so we regard CDCC as the exact calculation in this study.

Figure \ref{fig1} shows the $q$ dependence of $d\sigma/d\Omega$ 
for $d$ scattering from (a)  $^{12}$C target  
and (b) $^{208}$Pb target at the incident energies 
$E_{\rm in}/A_{\rP} = 20$--$200$ MeV. 
The dashed and dotted lines stand for the results of the DSF model
and CDCC, respectively. The result of the EP model is
shown by the solid line.
The difference between the dashed and dotted lines represents breakup
effects of $d$ on the elastic scattering that become more important at
forward angles as the incident energy decreases.
One sees that the EP model well reproduces the results of CDCC for both
$^{12}$C and $^{208}$Pb targets for $E_{\rm in}/A_{\rP}\geq80$ MeV. 
For $E_{\rm in}/A_{\rP}\leq40$ MeV,
agreement between the EP model and CDCC is reasonably well although
there is rather different behavior for the oscillation. 

\begin{figure*}[htbp]
 \begin{center}
  \includegraphics[width=0.42\textwidth,clip]{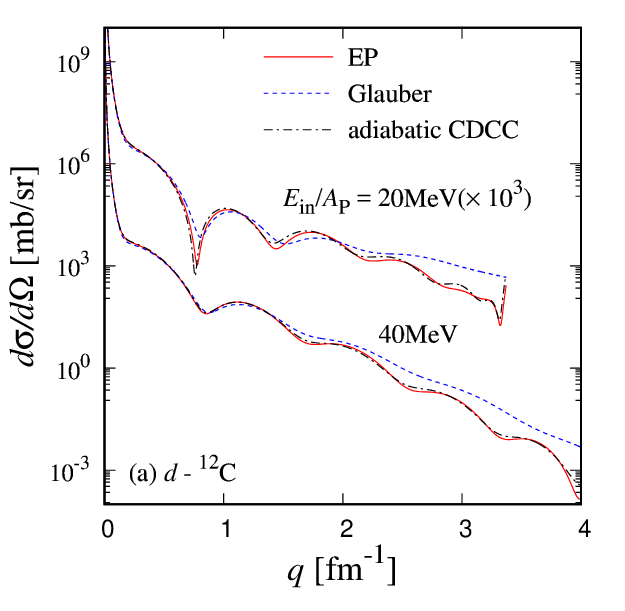}
  \includegraphics[width=0.42\textwidth,clip]{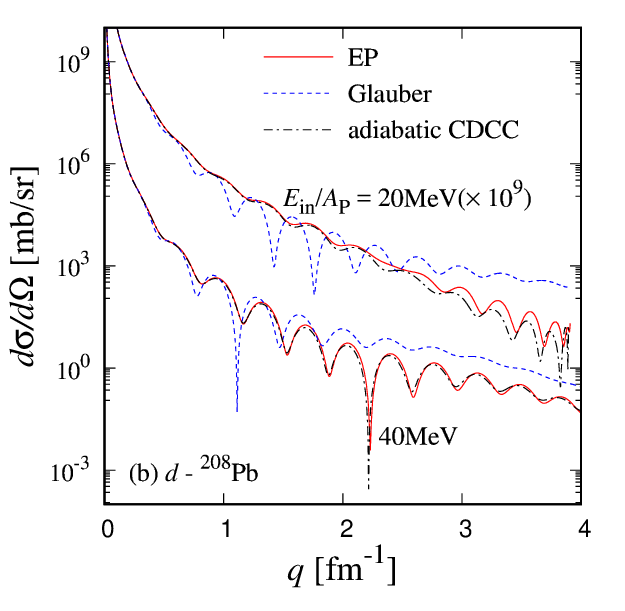}
  \caption{
  Differential cross sections $d\sigma/d\Omega$ as a function of 
  transfer momentum $q$ for $d$ scattering from (a) $^{12}$C 
  and (b) $^{208}$Pb target at $E_{\rm in}/A_{\rP} = 20$--$40$ MeV. 
  The solid line represents the result of the EP model, 
  the dashed line denotes the result of the Glauber model, 
  and the dot-dashed line corresponds to the result of adiabatic-CDCC,
  respectively. 
  Each cross section is multiplied by the factor shown in the figure. }
  \label{fig2}
 \end{center}
\end{figure*}

Here we consider the cause of the discrepancy between the EP model and
CDCC for $E_{\rm in}/A_{\rP}\leq40$ MeV.
The EP model includes the inaccuracy of the eikonal and adiabatic 
approximations because the EP model is based on the Glauber model. 
From the definition of the EP model, if we apply the eikonal
approximation to the EP model, the obtained eikonal $S$ matrix becomes
the same as ${\cal S}_{\rm N}$ of the Glauber model.
This means that the EP model might be regarded as the Glauber 
model without the eikonal approximation, i.e. the calculation with only 
the adiabatic approximation.
To clear this point, we calculate the differential cross section for $d$
scattering by the Glauber model and CDCC with the adiabatic
approximation (adiabatic-CDCC). In adiabatic-CDCC, energies for all
excited states are replaced by the ground-state energy.   

Figure~\ref{fig2} shows the results of the EP model, the Glauber model, and
adiabatic-CDCC for $d$ scattering from the $^{12}$C target (a) and
the $^{208}$Pb target (b) at $E_{\rm in}/A_{\rP}\leq$  40 MeV.  The solid,
dashed, and dot-dashed lines represent the results of the EP model, the
Glauber model, and adiabatic-CDCC, respectively. 
One sees that the difference between the Glauber model and
adiabatic-CDCC is not negligible. This means that the eikonal approximation 
is inaccurate because of the low-energy scattering. 
In particular, the results of the Glauber 
model are significantly different from those of adiabatic-CDCC
for $d$ scattering from $^{208}$Pb. The
reason is that the Coulomb interaction cannot be treated precisely in
the eikonal approximation in addition to the low-energy scattering.  
On the other hand, the EP model simulates well the results of adiabatic-CDCC.
This result shows that the discrepancy between the EP model and CDCC shown in 
Fig.~\ref{fig1} represents mainly the inaccuracy of the adiabatic
approximation. Thus we conclude that the inaccuracy of the eikonal 
approximation in the EP model is partially excluded.

\begin{figure}[htbp]
\begin{center}
 \includegraphics[width=0.41\textwidth,clip]{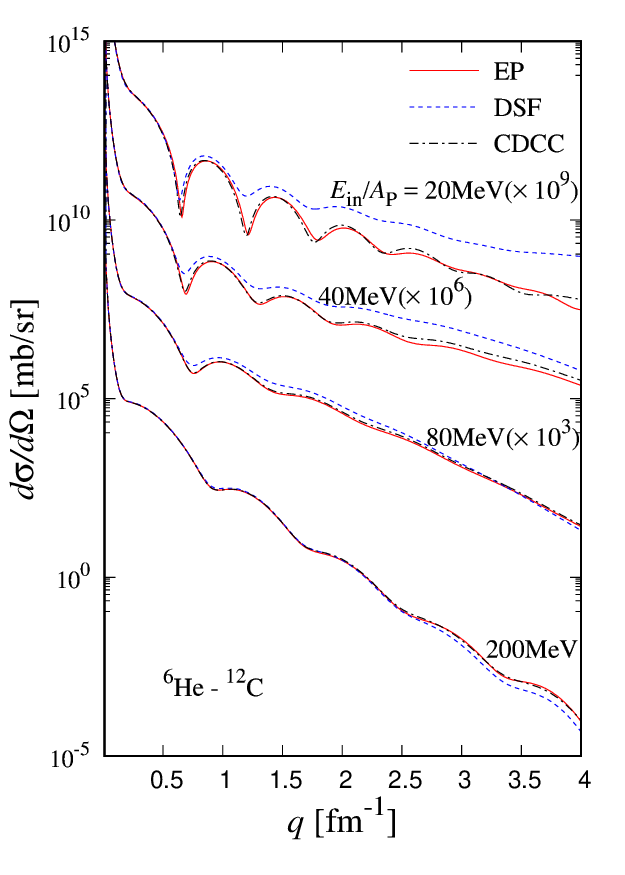}
\caption{
 The same as Fig.~\ref{fig1}, but for $^6$He scattering from $^{12}$C 
  at $E_{\rm in}/A_{\rP} = 20$--$200$ MeV. 
}
\label{6He12C-cs}
\end{center}
\end{figure}

\begin{figure}[htbp]
\begin{center}
 \includegraphics[width=0.6\textwidth,clip]{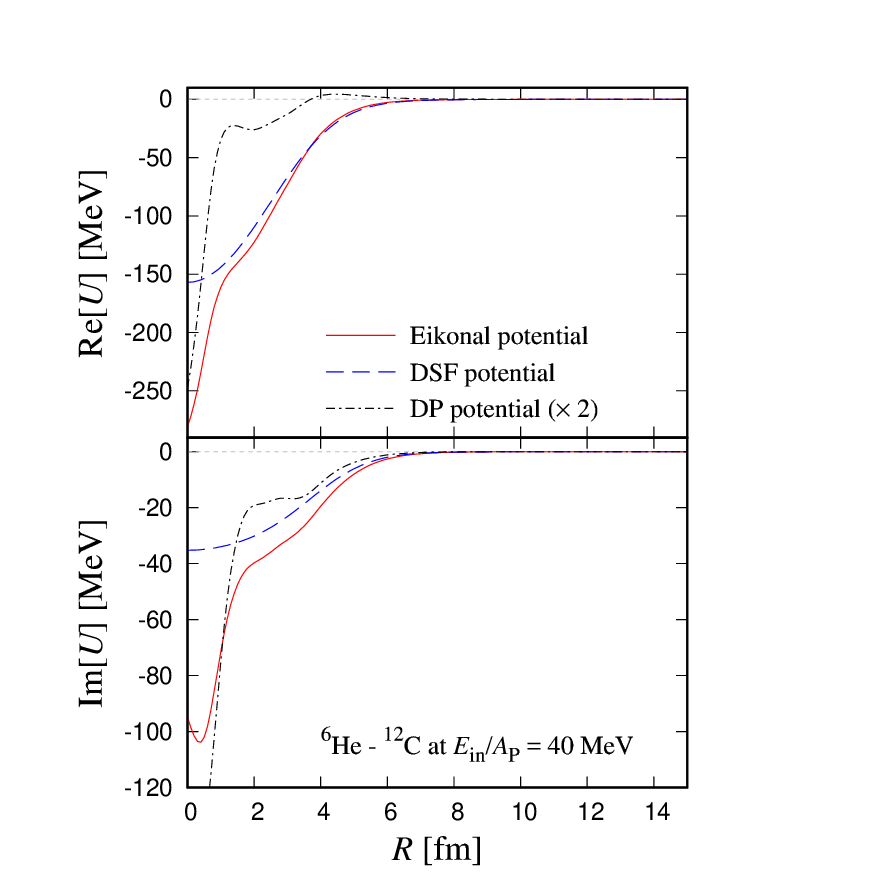}
\caption{
The $R$ dependence of the potentials for $^6$He scattering from
$^{12}$C at $E_{\rm in}/A_{\rP}=40$ MeV. The solid and dashed lines
 represent the eikonal potential and the DSF potential. The dot-dashed
 line stands for the dynamical polarization potential and is multiplied
 by 2. 
}
\label{6He12C-DPP}
\end{center}
\end{figure}

\begin{figure}[htbp]
\begin{center}
 \includegraphics[width=0.41\textwidth,clip]{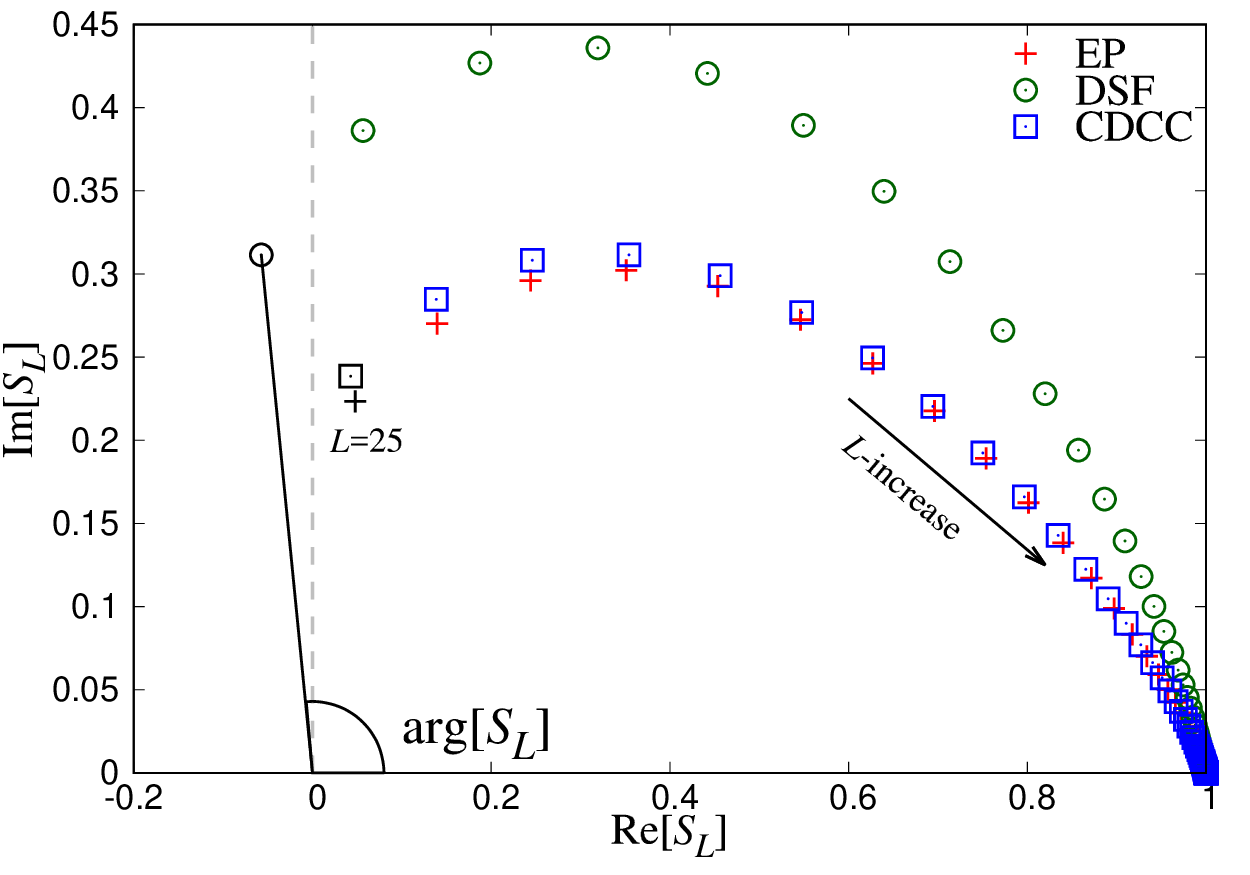}
\caption{
  The $S$ matrices for $^6$He scattering from  $^{12}$C at 
 $E_{\rm in}/A_{\rP}=40$ MeV
 are plotted from the grazing angular momentum.
 The cross, circle, square points describe the results
 of the EP model, the DSF model CDCC, respectively.
 $L$ means the orbital angular momentum,  and increases toward the
 $S=(1,0)$. 
}
\label{6He12C-Smat}
\end{center}
\end{figure}

\subsection{differential cross section of $^6$He scattering}
\label{Result:ep-smat}
Next we apply the EP model to analyses of $^6$He + $^{12}$C scattering,
which is described as a $^4$He + $n$ + $n$ + T four-body model.
As the exact calculation for the four-body scattering system,
we adopt four-body CDCC that has been successful in
analyses of breakup reaction of $^6$He~\cite{Mat04,Mat06,Rod08}.
In the EP model for the four-body system, the eikonal $S$ matrix defined
in Eq.~\eqref{eq:SN} is rewritten by 
\begin{eqnarray}
 {\cal S}_{\rm N}(\bm{b})=
  \langle\Phi_0|e^{i\chi_{\rm N}(\mbold{\xi},\bm{b})}|\Phi_0\rangle,
  \label{eq:SN4} 
\end{eqnarray}
and the phase-shift function is defined as
\begin{eqnarray}
 \chi_{\rm N}(\mbold{\xi},\bm{b})
  &=&-\frac{1}{\hbar v}
  \int_{-\infty}^{\infty}U(\mbold{\xi},\mbold{R})dZ
  \label{eq:chiN4}
\end{eqnarray}
with
\begin{eqnarray}
 U(\mbold{\xi},\mbold{R})&=&U_{n}(\mbold{\xi},\mbold{R})
  +U_{n}(\mbold{\xi},\mbold{R})+U_\alpha(\mbold{\xi},\mbold{R}).
\end{eqnarray}
Here $\mbold{\xi}$ represents the internal coordinate of $^6$He, and 
$U_n$ ($U_\alpha$) is the optical potential between $n$ ($^4$He) and T.
For $U_n$, we use the same potential used in the analysis of $d$
scattering, and $U_\alpha$ is obtained by folding $U_n$ and $U_p$ with
the $^4$He density. The ground state wave function of $^6$He, $\Phi_0$,
is calculated by the Gaussian expansion method (GEM) with a $^4$He + $n$
+ $n$ three-body model. In GEM, the model Hamiltonian of $^6$He and the
parameter set of the Gaussian basis functions are adopted the same ones
used in Ref.~\cite{Kik13}. 

Figure~\ref{6He12C-cs} shows the results of the EP model (the solid
line), the DSF model (the dotted line), 
and CDCC (the dot-dashed line) for the $^6$He scattering from $^{12}$C at 
$E_{\rm in}/A_{\rm P}=20$--200 MeV. 
One sees that breakup effects of
$^6$He represented by the difference between the results of the DSF
model and CDCC are stronger than those of $d$. Furthermore the EP model
well reproduces the results of CDCC even at $E_{\rm in}/A_{\rm P}\leq40$
MeV.
One of the reason is that the condition represented by 
Eq.~\eqref{eq:Eik-cond} is satisfied well
because $K$ of the $^6$He scattering is larger 
than $K$ of the $d$ scattering in the same $E_{\rm in}/A_{\rm P}$.
Thus we expect that the EP model works well for
scattering with heavier projectiles. 

\subsection{dynamical polarization potential}

Finally, we discuss breakup effects of the optical potential for the
$^6$He scattering. Figure~\ref{6He12C-DPP} shows the optical potentials
derived by the EP model and the DSF model for $^6$He scattering from
$^{12}$C at $E_{\rm in}/A_{\rP} = 40$ MeV. 
The solid line represents the eikonal potential, and the dashed line
corresponds to the DSF potential that doesn't include breakup effects of
$^6$He.  
The dot-dashed line stands for the dynamical polarization (DP) potential 
$U_{\rm DP}$
defined as 
\begin{eqnarray}
 U_{\rm DP}(R)&=&U_{\rm E}(R)-U_{\rm DSF}(R),
\end{eqnarray}
which represents breakup effects of $^6$He on the optical potential 
between $^6$He and $^{12}$C.
In this figure, the strength of the DP potential is multiplied by 2.
In Fig.~\ref{6He12C-DPP}, the real part of the DP potential is repulsive
in the peripheral region and the imaginary part is absorptive in the whole region.
This behavior of the DP potential can be understood from the $S$
matrices for the scattering. In the eikonal approximation as shown in
Eqs.~\eqref{eq:SN4} and \eqref{eq:chiN4}, the real part of the optical
potential is related to the argument of the $S$ matrix, and the
imaginary part of the optical potential is the absolute value of the $S$
matrix.  
Figure~\ref{6He12C-Smat} shown the $S$ matrices for the scattering. The
cross, circle and square marks stand for the results of the EP model,
the DSF model, and CDCC, respectively. The $S$ matrices are plotted from
the grazing angular momentum $L=25$, which is defined this momentum as a
position of the peak at the partial breakup cross section. 
The difference between the results of the DSF model and CDCC represents
breakup effects on the $S$ matrix, and the results of the EP model are
in good agreement with those of CDCC. One sees that the absolute value
of the $S$ matrix for CDCC is smaller than that for the DSF model, and
the argument of the $S$ matrix for CDCC is also smaller than that for
the DSF model. This result shows that breakup effects make weak for the
real part and strong for the imaginary part of the optical potential.

\section{Summary} 
\label{Summary} 

We construct a microscopic optical potential including breakup effects
based on the Glauber model, in which a NA
potential is derived by the folding model with
the Melbourne $g$ matrix. The microscopic optical potential is referred 
to as the eikonal potential. 

First, in order to confirm the validity of the eikonal potential, 
we compared the EP model with CDCC for the elastic cross
sections for $d$ scattering from $^{12}$C and $^{208}$Pb at
$E_{\rm in}/A_{\rP}=20\mbox{--}200$ MeV, where the scattering system 
is described by the $p$ + $n$ + T three-body model.
As the result, the EP model well reproduces the results of CDCC as the
incident energy increases.
In the analysis, we found that the difference between the results of the EP
model and CDCC at low incident energies comes from mainly the inaccuracy 
of the adiabatic approximation, and the inaccuracy of the eikonal approximation
is partially excluded from the EP model. In fact, the EP model simulates well
the result of adiabatic-CDCC for $E_{\rm in}/A_{\rP}\leq40$ MeV.

Next, we apply the EP model to analyses of $^6$He scattering from $^{12}$C
at $E_{\rm in}/A_{\rP}=20\mbox{--}200$ MeV, where the scattering system is 
described by a $^4$He + $n$ + $n$ + $^{12}$C four-body model.
In the analyses of $^6$He scattering, the EP model well reproduces the 
results of CDCC even at $E_{\rm in}/A_{\rm P}\leq$40 MeV.
This means that because the wave number of the $^6$He
scattering is larger than that of the $d$ scattering in the same 
$E_{\rm in}/A_{\rm P}$, the eikonal approximation is valid even at 
$E_{\rm in}/A_{\rm P}\leq$40 MeV. Thus the EP model works well 
for heavier projectile scattering. Furthermore we investigated breakup 
effects on the optical potential via the dynamical polarization potential. 

One of the advantage of the EP model is to be applicable to analyses for
scattering of many-body projectiles as well as the Glauber model.
In the forthcoming paper, we will try to apply the EP model to analyses of
scattering of $^8$He, in which $^8$He is described by a $^4$He + 
$n$ + $n$ + $n$ + $n$ five-body model. Furthermore, we will discuss Coulomb
breakup effects, which are omitted in the present calculation. The
treatment of Coulomb breakup processes within the eikonal approximation
has been discussed in some papers~\cite{Oga03,Cap08}, and the problem
is an important subject for 
the eikonal calculation. 

\section*{Acknowledgements} 
This work is supported in part by
by Grant-in-Aid for Scientific Research
(Nos. 25400266, and 26400278)
from Japan Society for the Promotion of Science (JSPS).



\end{document}